\documentclass[11pt]{article}
\usepackage{amssymb,amsmath,amsfonts}
\usepackage{graphicx}
\usepackage{graphics}
\usepackage{eepic,epsfig}

\begin{document}
	
	\title{Induced current by a cosmic string and a brane in high-dimensional AdS spacetime}
	\author{W. Oliveira dos Santos$^{1}$%
		\thanks{%
			E-mail: wagner.physics@gmail.com} ,\thinspace\ E. R. Bezerra de Mello$^{1}$\thanks{%
			E-mail: emello@fisica.ufpb.br} \\
		\\
		$^{1}$\textit{Departamento de F\'{\i}sica, Universidade Federal da Para\'{\i}%
			ba}\\
		\textit{58.059-970, Caixa Postal 5.008, Jo\~{a}o Pessoa, PB, Brazil}}
	\maketitle
	
	\begin{abstract}
	In this paper we investigate the bosonic current induced by a brane and 
	a magnetic flux running along the idealized cosmic string in a $(D+1)$-dimensional
	anti-de Sitter (AdS) background. We consider the brane is parallel to the AdS boundary
	and the cosmic string is orthogonal to them. Moreover, we assume that on the 
	brane the charged bosonic field obeys the Robin boundary condition. 
	The brane divides the space into two regions with different
	properties of the vacuum state. We show that the only nonzero component of the current
	density is along the azimuthal direction in both regions. In order to develop this analysis
	we calculate, for both regions, the positive frequency Wightman functions. Both functions present a part associated
	with the AdS in presence of a cosmic string only, and the other part induced by the brane. In this paper we consider only the contributions induced by the brane. We show that in both regions the azimuthal current densities are odd functions of the magnetic flux along the string. Different analytic and numerical analysis are performed and an application of our results is provided for the Randall-Sundrum braneworld model with a single brane.
	\end{abstract}
	
	\bigskip
	
	
	\bigskip
	
	\section{Introduction}
	
One of the most fascinate topological object predicted by the Grand Unified Theory as consequence of gauge symmetry breaking is the cosmic string \cite{Kibble,V-S}. Although recent observational data on the cosmic microwave background have discarded cosmic strings as the primary source for primordial density	perturbation, these objects are still candidate for the generation of a number of	interesting physical effects such as gamma ray bursts \cite{Berezinski},	gravitational waves \cite{Damour} and high energy cosmic rays \cite	{Bhattacharjee}. The gravitational field produced by a cosmic string may be approximated by a planar angle deficit in the two-dimensional sub-space. The strength of gravitational interactions of cosmic strings with matter is its {\it tension}, that is characterized by the dimensionless parameter, $G\mu_0$, defined in natural units. In this expression $G$ represents the Newton's gravitation constant and $\mu_0$ the linear mass density of the string, which is proportional to the square of the energy scale where the gauge symmetry is broken.
	 
The anti-de Sitter	(AdS) spacetime is a solution of the Einstein equation in presence of negative cosmological constant. Being maximally symmetric spacetime, it allowed us to solve many problems in quantum fields exactly (see, for example \cite{Fronsdal}-\cite{Caldareli}).  Besides, the importance of this background has increased when it was observed that it generically arises as a ground state in extended supergravity and in string theories. Moreover, additional interest in this geometry was generated by the appearance of two models where AdS spacetime plays a special role. The first model, the AdS/CFT correspondence (for a review see \cite{Ahar00}), represents a realization of the holographic principle and relates string theories or supergravity in the AdS bulk with a conformal field theory 	living on its boundary. The second model is the braneworld scenario with large extra dimensions. This model offers a solution to the hierarchy energy scale problem associated with the gravitational and electroweak interactions (for comprehensive discussions on braneworld gravity and cosmology, see \cite{Brax03,Maar10}).

The analysis of the spacetime geometry due to a cosmic string in AdS bulk has been considered in \cite{Ghe1,Cristine}. There it was shown that at distances larger than the string's core radius, the gravitational effects due to the presence of the string is well described by a planar deficit angle in the AdS metric, similarly to the case in the Minkowskian bulk. This non-trivial topology by its turn provides additional vacuum polarization effects. In this way the combined effect of the curvature and non-trivial topology contribute to the evaluation of the vacuum expectation value of several physical observables, as the energy-momentum tensor.  
	
The investigation of the vacuum expectation value (VEV) of the bosonic current density, $\langle j^\mu\rangle$, and the energy-momentum tensor, $\langle T^\mu_\nu\rangle$, induced by an  idealized cosmic string carrying magnetic flux running along its core in a $(D+1)-$dimensional  AdS spacetime, have been analyzed in \cite{Wagner_19} and \cite{Wagner_20}, respectively. Moreover, in both papers it was admitted a compactification of one dimension along the string, and the presence of an extra magnetic flux running its center. Also the study of the VEV of fermionic energy-momentum tensor and current density in a $(1+4)-$dimensional AdS space with a compactified cosmic string have been considered in \cite{Wagner_20a} and \cite{Wagner_22}, respectively.  The analysis of the effects due to a brane on the vacuum fermionic current, $\langle j^\mu\rangle$, and the energy-momentum tensor, $\langle T^\mu_\nu\rangle$, with the brane parallel to the AdS boundary, were studied in \cite{Wagner_21} and \cite{Wagner_22a}, respectively. The analysis of the VEV of the energy-momentum tensor associated with a charged bosonic field on the AdS background in the presence of a cosmic string considering  a brane parallel to the AdS boundary, was developed in \cite{Wagner_23}. Here in this paper, we want to continue in the same line of investigation, but at this time we will turn our attention to analyze the effects of the brane on the VEV of the induced current in both regions defined by the brane. 
	
The organization of the paper is the following: In section \ref{sec2} we present the geometry of the spacetime that we want to consider, and the complete set of normalized positive energy solutions of the Klein-Gordon equation considering the presence of a brane parallel to the AdS boundary. In section \ref{Wight_fun} we construct the Wightman functions for both regions of the space separated by the brane. These regions are: between the AdS boundary and the brane ($L$-region) and between the brane and AdS horizon ($R$- region). The corresponding Wightman functions are decomposed in a part due to the AdS spacetime in presence of a cosmic sting in the absence of brane, plus the ones induced by the brane. In section \ref{bosonic_cur} we evaluate the VEV of the bosonic current densities in both regions.  Because the above mentioned decomposition of the Wightman functions, the same happens for the current densities. Also in this section, various asymptotic limits of the currents are considered and numerical results are presented. In Section \ref{RS} we apply our analysis to the Randal-Sundrum type model with a single brane. In Section \ref{conc} we summarize the most relevant result obtained. Throughout the paper, we use natural units $G=\hbar =c=1$.

\section{Klein-Gordon equation}
	\label{sec2}
The main objective of this section is to provide the complete set of normalized solutions of the Klein-Gordon equation associated with a massive scalar charged quantum field propagating in a $(D+1)$-dimensional AdS spacetime, with $D\geq 3$, in presence of a magnetic-flux-carrying cosmic string and taking into account the presence of a brane parallel to the AdS boundary.  So, with this objective  we present first, the line element, in cylindrical coordinate, in $(1+3)-$dimensional AdS spacetime in the presence of a cosmic string:
\begin{equation}
		ds^{2}=e^{-2y/a}[dt^{2}-dr^{2}-r^{2}d\phi ^{2}]-dy^{2}\ .  \label{ds1}
\end{equation}
In the above line element the idealized cosmic string is along the $y-$axis,  $r\geqslant 0$ and $\phi \in \lbrack 0,\ 2\pi /q]$ define the coordinates on the conical geometry, $(t, \ y)\in (-\infty ,\ \infty )$. The parameter $a$ is associated with the curvature scale of the background spacetime; moreover, the parameter $q> 1$ provides the planar angle deficit, $\delta\phi=2\pi(1-q^{-1})$, produced by the cosmic string. In our analysis, we will use the \textit{Poincar\'{e}} coordinate defined by $w=ae^{y/a}$. In this case the line element above is expressed in the form conformally related to the line element associated with a cosmic string in Minkowski spacetime:
\begin{equation}
	ds^2 = \left(\frac{a}{w}\right)^2[dt^2 - dr^2 - r^2d\phi^2 - dw^2 ]  \  . 
	\label{ds2}
\end{equation}
The new coordinate, $w$, is defined in the interval $\lbrack 0,\ \infty )$. Two values for this coordinates deserve to be mentioned: $w=0$ and $w=\infty $. They correspond to the AdS boundary and horizon, respectively.  The generalization of (\ref{ds2}) to $(D+1)$-dimensional, with $D>3$, is given by
\begin{equation}
	\label{ds2_gen}
	ds^2 = \left(\frac{a}{w}\right)^2[dt^2 - dr^2 - r^2d\varphi^2 - dw^2 - \sum_{i=4}^{D}(dx^i)^2]   \   ,
\end{equation}
with $x^i\in (-\infty ,\ \infty )$. In the above generalized metric tensor, the defect that creates the conical structure is not a linear one; in fact, for this case, there is a string-like structure with $D-2$ dimensions.

The curvature scale $a$ in \eqref{ds2_gen} is related to the cosmological constant, $\Lambda $, and the Ricci scalar, $R$, by the formulas
\begin{equation}
	\Lambda =-\frac{D(D-1)}{2a^{2}} \ ,\ \ R=-\frac{D(D+1)}{a^{2}}\ .
	\label{LamR}
\end{equation}

\subsection{Klein-Gordon equation}
The field equation which governs the quantum dynamics of a charged
bosonic field with mass $m$, in a curved background and in the presence of an 
electromagnetic potential vector, $A_\mu$, reads
	\begin{equation}
		({\mathcal{D}}^2 + m^2 + \xi R)\varphi(x) = 0  \   , 
		\label{KGE}
	\end{equation}
	where the differential operator in the field equation reads
	\begin{align}
		{\mathcal{D}}^2=\frac{1}{\sqrt{|g|}}{\mathcal{D}}_{\mu}\left(\sqrt{|g|}g^{\mu \nu} {\mathcal{D}}_{\nu
		}\right), \ {\mathcal{D}}_{\mu
		}=\partial _{\mu }+ieA_{\mu }\   \ {\rm with} \  \  g=\det(g_{\mu\nu})  \  .  \label{1}
	\end{align}
In the above equation, we also consider the presence of a non-minimal coupling, $\xi$, between the field and the geometry represented by the Ricci scalar, $R$. Two specific values for the curvature coupling are of special interest: the value $\xi = 0$ corresponds to minimal coupling, and and $\xi = \frac{D - 1}{4D}$, the conformal coupling, respectively. Moreover, we consider only the component vector potential $A_{\phi}=-q\Phi_\phi/(2\pi)$ different from zero. This vector potential corresponds to a magnetic flux, $\Phi_\phi$, running along the string's axis.
	
In this work we admit the presence of a codimension one flat boundary, hereafter named brane, located at $w=w_0$ and parallel to the AdS boundary, and that the the field operator obeys the gauge invariant Robin boundary condition,
	\begin{equation}
		(1+\beta n^{\mu}{\mathcal{D}}_{\mu
		})\varphi(x)=0, \ w=w_0 \ .
		\label{RBC}
	\end{equation}
The parameter $\beta$ in the equation above is a constant and it encodes the properties of the brane. In particular, the special cases $\beta=0$ and $\beta=\infty$ correspond to the Dirichlet and Neumann boundary conditions, respectively. In addition, $n^{\mu}$ is the inward pointing (with respect to the region under consideration) normal to the brane at $w=w_0$. In the region $0\le w\le w_0$ (referred to as the $L$(left)-region this normal is given as $n^{\mu}=-\delta_{3}^{\mu}a/w$ and in the region $w_0\le w\le \infty$ $R$(right)-region as  $n^{\mu}=\delta_{3}^{\mu}a/w$.
	
In the geometry defined by \eqref{ds2_gen} and in the presence of the azimuthal vector potentials, $A_\phi$, the Klein-Gordon (KG) equation \eqref{KGE} becomes
	\begin{eqnarray}
		\left[\frac{\partial^2}{\partial t^2} - \frac{\partial^2}{\partial r^2} - \frac{1}{r}\frac{\partial}{\partial r} - \frac{1}{r^2}\left(\frac{\partial}{\partial\phi} + ieA_{\phi}\right)^2
		- \frac{\partial^2}{\partial w^2}-\frac{(1-D)}{w}\frac{\partial}{\partial w}
		\right.\nonumber\\
		\left. + \frac{M(D,m,\xi)}{w^2} - \sum_{i=4}^{D}\frac{\partial^2}{\partial (x^i)^2} \right]\varphi(x) = 0  \  , 
		\label{KGE2}
	\end{eqnarray}
with $M(D,m,\xi) = a^2m^2 - \xi D(D+1)$. 
	
On basis of previous analysis presented in \cite{Wagner_19}, the positive energy solution of \eqref{KGE2} can be expressed by
	\begin{equation}
		\varphi_{\sigma}(x) = C_{\sigma}w^{\frac{D}{2}}W_{\nu}(pw)J_{q|n +\alpha|}(\lambda r)e^{- iE t + iqn\phi + i\vec{k}\cdot\vec{x}_{\parallel}} \ ,
		\label{Solu1}
	\end{equation}
with the function $W_\nu(pw)$,
	\begin{equation}
		W_{\nu}(pw)=C_1J_{\nu}(pw)+C_2Y_{\nu}(pw) \  ,
		\label{W-function}
	\end{equation}
being given by a linear combination of the Bessel and Neumann functions \cite{Abra}. The order of these functions is,
	\begin{equation}
		\nu = \sqrt{\frac{D^2}{4} + a^2m^2 - \xi D(D+1)} \ .
		\label{nu}
	\end{equation}
The enrgy and the parameter $\alpha$ in the Bessl function associated with the radial coordinate are:
	\begin{eqnarray}
		E &=& \sqrt{\lambda^2 + p^2 + \vec{k}^2},\nonumber\\
		\alpha &=& \frac{eA_{\phi}}{q} = -\frac{\Phi_{\phi}}{\Phi_0} \ , 
		\label{const}
	\end{eqnarray}
	being $\Phi_0=\frac{2\pi}{e}$, the quantum flux. Moreover, in \eqref{Solu1} $\vec{x}_{\parallel}$ corresponds the coordinates defined in the $(D-3)$ extra dimensions, and $\vec{k}$ the corresponding momentum. In \eqref{Solu1},  $\sigma$ represents the set of quantum numbers $(n, \lambda, p, \vec{k})$, with $n=0,\pm1,\pm2,\ldots$, $\lambda \geq 0$, $-\infty<k^j<\infty$ for $j=4,...,D$. As to the quantum number $p$, it is determined, separately, for each region divided by the brane.
	
The coefficient $C_{\sigma}$ in \eqref{Solu1} is determined from the normalization condition
	\begin{eqnarray}
		\int d^Dx\sqrt{|g|}g^{00}\varphi_{\sigma'}^{*}(x)\varphi_{\sigma}(x)= \frac{1}{2E}\delta_{\sigma,\sigma'}  \   ,
		\label{NC}
	\end{eqnarray}
with delta symbol on the right-hand side is understood as Dirac delta function for the continuous quantum number, and Kronecker delta for the discrete one.

\subsubsection{Normalized wave-functions in $R$-region}	
First let us consider the $R$-region. Due to the Robin boundary condition \eqref{RBC} on the flat boundary, we obtain the relation $C_2/C_1=-\bar{J}_{\nu}(pw_0)/\bar{Y}_{\nu}(pw_0)$ for the coefficients in \eqref{W-function}. 

From now on, we use the compact notation
\begin{equation}
	\label{F_function}
		\bar{F}(x)=A_{0}F(x)+B_{0}xF^{\prime}(x) \ ,
\end{equation}
with the coefficients
	\begin{equation}
		\label{coef_1}
		A_{0}=1+\frac{D\beta}{2a}, \quad B_{0}=\beta/a \ , 
	\end{equation}
in the expressions of the wave-functions. 

The normalized solutions of KG equation in $R$-region compatible with the boundary condition \eqref{RBC}, can be written presented as,
	\begin{equation}
		\varphi_{(R)\sigma}(x) = C_{(R)\sigma}w^{\frac{D}{2}}g_{\nu}(pw_0,pw)J_{q|n +\alpha|}(\lambda r)e^{- iE t + iqn\phi + i\vec{k}\cdot\vec{x}_{\parallel}} \ ,
		\label{Solu-R}
	\end{equation}
	where we have introduced the function
	\begin{equation}
		\label{g-function}
		g_{\nu}(u,v)=J_{\nu}(v)\bar{Y}_{\nu}(u)-\bar{J}_{\nu}(u)Y_{\nu}(v) \ .
	\end{equation}
Due to the  continuous values assumed by the quantum number $p$, the normalization condition \eqref{NC}, provides
	\begin{equation}
		|C_{(R)\sigma}|^2=\frac{(2\pi)^{2-D}qp\lambda}{2Ea^{D-1}[\bar{J}_{\nu}^2(pw_0)+\bar{Y}_{\nu}^2(pw_0)]} \ .
		\label{C_R-region}
	\end{equation}

\subsubsection{Normalized wave-functions in $L$-region}
	
In the $L$-region, the integration over $w$ in \eqref{NC}, is restricted in the interval $0\le w\le w_0$. In the case of $C_2\neq0$ in \eqref{W-function}, the integral over $w$ diverges at the lower limit $w=0$ for the case with $\nu\ge1$. Consequently, in that region, we should take $C_2=0$. On the other hand, in the interval $0\le \nu <1$, the solution \eqref{W-function} with $C_2\neq0$ is normalizable and in order to uniquely define the mode functions an additional boundary condition at the AdS boundary	is required \cite{Breitenlohner,Avis1978}. Here, we assume the Dirichlet boundary condition on $w=0$ which implies $C_2=0$. Thus, with this choice, the mode function in the $L$-region are given by
	\begin{equation}
		\varphi_{(L)\sigma}(x) = C_{(L)\sigma}w^{\frac{D}{2}}J_{\nu}(pw)J_{q|n +\alpha|}(\lambda r)e^{- iE t + iqn\phi + i\vec{k}\cdot\vec{x}_{\parallel}} \ .
		\label{Solu-L}
	\end{equation}
According to the Robin boundary condition \eqref{RBC}, the eigenvalues of the quantum number $p$ obey the relation:
	\begin{equation}
		\label{J_barr}
		\bar{J}_{\nu}(pw_0)=0 \ ,
	\end{equation}
being $	\bar{J}_{\nu}(x)$  given by \eqref{F_function}, with
	\begin{equation}
	\label{coef_1}
	A_{0}=1-\frac{D\beta}{2a}, \quad B_{0}=-\beta/a \ .
\end{equation}
The eigenvalues of \eqref{J_barr} are given by $p=p_{\nu,i}/w_0$, with $p_{\nu,i}$ being the positive zeros of the function $\bar{J}_{\nu}(x)$, enumerated by $i=1, 2,...$. We can observe that the roots $p_{\nu,i}$ do not depend on the location of the brane. Considering now the normalization condition \eqref{NC}, with $\delta_{p,p^\prime}=\delta_{i,i^\prime}$, and integrating over $w$ in the interval $[0,w_0]$, after some algebraic manipulations involving the Bessel functions, we obtain
	\begin{equation}
		|C_{(L)\sigma}|^2=\frac{(2\pi)^{2-D}qp_{\nu,i}\lambda T_{\nu}(p_{\nu,i})}{w_0a^{D-1}\sqrt{p_{\nu,i}^2+w_{0}^2(\lambda^2+\vec{k}^2})}\ .
		\label{C_L-region}
	\end{equation}
In the above expression we have for the function $T_\nu(z)$, the following result:
\begin{eqnarray}
T_{\nu}(z)=z[(z^2-\nu^2)J_{\nu}^2(z)+z^2(J_{\nu}^{\prime}(z))^2]^{-1} \ .
\end{eqnarray} 
	
	\section{Wightman Function}
	\label{Wight_fun}
	In this section we want to obtain the positive frequency Wightman function induced by the brane for both regions, $R$ and $L$, in a closed form. The positive frequency Wightman function is defined by  $W(x,x^{\prime})=\langle 0|\hat{\varphi}(x)\hat{\varphi}^{\dagger}(x^{\prime})|0\rangle$, where $|0\rangle$ stands for the vacuum state. Here we assume that the field operator, $\hat{\varphi}(x)$, is prepared in the Poincaré vacuum state. To evaluate this function, we use the mode sum formula
	\begin{equation}
		W(x,x^{\prime})=\sum_{\sigma}\varphi_{\sigma}(x)\varphi_{\sigma}^{\ast}(x^{\prime}) \ .
		\label{Wightman-def}
	\end{equation}
\subsection{$R$-region}
The normalized positive energy solution of the KG equation in $R$-region, is given by combining previous results, \eqref{Solu-R}, \eqref{g-function} and \eqref{C_R-region}. It reads,
\begin{equation}
	\varphi_{(R)\sigma}(x) = \sqrt{\frac{(2\pi)^{2-D}qp\lambda}{2Ea^{D-1}[\bar{J}_{\nu}^2(pw_0)+\bar{Y}_{\nu}^2(pw_0)]}}w^{\frac{D}{2}}g_{\nu}(pw_0,pw)J_{q|n +\alpha|}(\lambda r)e^{- iE t + iqn\phi + i\vec{k}\cdot\vec{x}_{\parallel}} \ .
	\label{Solu-R_total}
\end{equation}

Substituting the above expression into \eqref{Wightman-def}, we get:
\begin{eqnarray}
	W_R(x,x^{\prime})&=&\frac{q(ww')^{D/2}}{2a^{D-1}(2\pi)^{D-2}}\sum_\sigma\frac{p\lambda}E\frac{g_{\nu}(pw_0,pw)g_{\nu}(pw_0,pw^\prime)}{(\bar{J}_{\nu}^2(pw_0)+\bar{Y}_{\nu}^2(pw_0))}J_{q|n +\alpha|}(\lambda r)J_{q|n +\alpha|}(\lambda r^\prime)\nonumber\\&\times&e^{- iE(t-t^\prime) + iqn(\phi-\phi^\prime) + i\vec{k}\cdot(\vec{x}_{\parallel}- \vec{x}^\prime_{\parallel})} \  ,  \label{W_R}
\end{eqnarray}
where we are using the compact notation below for the summation over $\sigma$:
\begin{equation}
	\sum_{\sigma }=\sum_{n=-\infty}^{+\infty} \ \int dp \ \int_0^\infty
	\ d\lambda \ \int d{\vec{k}} \ .  \label{Sumsig}
\end{equation}

Now performing a Wick rotation on the time coordinate and using the identity
\begin{equation}
	\frac{e^{-E\Delta \tau}}{E}=\frac2{\sqrt{\pi}}\int_0^\infty ds e^{-s^2E^2-(\Delta \tau)^2/(4s^2)}  \  ,
	\label{identity}
\end{equation}
being $E=\sqrt{\lambda^2+p^2+\vec{k}^2}$, we can integrate over $\lambda$ and $\vec{k}$ with the help of \cite{Grad}. The final result result is
\begin{eqnarray}
	W_R(x,x^\prime)&=&\frac{qrr^\prime}{2(2\pi)^{D/2}a^{D-1}}\left(\frac{ww^\prime}{rr^\prime}\right)^{D/2}\int_{0}^{\infty}dv v^{\frac{D}{2}-2}e^{-\frac{r^2+r^{\prime2}+\Delta\vec{x}_{\parallel}^2-\Delta t^2}{2rr^\prime}v}\sum_{n}e^{inq\Delta\phi}I_{q|n+\alpha|}(v)\nonumber\\
	&\times&\int_{0}^{\infty}dppe^{-\frac{rr^\prime}{2v}p^2}\frac{g_{\nu}(pw_0,pw)g_{\nu}(pw_0,pw^{\prime})}{\bar{J}_{\nu}^2(pw_0)+\bar{Y}_{\nu}^2(pw_0)} \ ,
	\label{W-function_R}
\end{eqnarray}
where we have introduced a new variable, $v=rr^\prime/(2s^2)$.

The above Wightman function contains the contributions coming from the cosmic string in AdS spacetime without boundary, more the contribution induced by the boundary. Because, in this paper, we are interested to calculate the VEV of the current induced by the presence of the brane, let us subtract form \eqref{W-function_R} the corresponding Wightman induced by the cosmic string only in AdS. The latter can be obtained from \eqref{W-function_R}, by taking the limit $w_0\to 0$. So we can write:
\begin{equation}
	W_{b(R)}(x,x^\prime)=W_R(x,x^\prime)-W_{cs}(x,x^\prime) \ .
	\label{Proc}
\end{equation}

The Wightman function, $W_{b}(x,x^\prime)$, is obtained by using the the following identity:
\begin{eqnarray}
\frac{g_{\nu}(pw_0,pw)g_{\nu}(pw_0,pw^{\prime})}{\bar{J}_{\nu}^2(pw_0)+\bar{Y}_{\nu}^2(pw_0)}-
	J_{\nu}(pw)J_{\nu}(pw^\prime)=-\frac{1}{2}\sum_{l=1}^{2}\frac{\bar{J}_{\nu}(pw_0)}{\bar{H}_{\nu}^{(l)}(pw_0)}H_{\nu}^{(l)}(pw)H_{\nu}^{(l)}(pw^\prime) \ ,
\end{eqnarray}
being $H_{\nu}^{(l)}(x)$, $l=1, 2$, the Hankel functions \cite{Abra}. So, we get,
\begin{eqnarray}
	W_{b(R)}(x,x^\prime)&=&-\frac{qrr^\prime}{4(2\pi)^{D/2}a^{D-1}}\left(\frac{ww^\prime}{rr^\prime}\right)^{D/2}\int_{0}^{\infty}dv v^{\frac{D}{2}-2}e^{-\frac{r^2+r^{\prime2}+\Delta\vec{x}_{\parallel}^2-\Delta t^2}{2rr^\prime}v}\sum_{n}e^{inq\Delta\phi}I_{q|n+\alpha|}(v)\nonumber\\
	&\times&\int_{0}^{\infty}dppe^{-\frac{rr^\prime}{2v}p^2}\sum_{l=1}^{2}\frac{\bar{J}_{\nu}(pw_0)}{\bar{H}_{\nu}^{(l)}(pw_0)}H_{\nu}^{(l)}(pw)H_{\nu}^{(l)}(pw^\prime) \ .
	\label{W-function_R_b}
\end{eqnarray}

The parameter $\alpha$ can be written in the form
\begin{equation}
	\alpha=n_{0}+\alpha_0, \ \textrm{with}\ |\alpha_0|<\frac{1}{2},
	\label{const-2}
\end{equation}
with $n_{0}$ being an integer number; moreover, the sum over the quantum number $n$ in \eqref{W-function_R_b}, has been developed in \cite{deMello:2014ksa}. The result is given below:
\begin{eqnarray}
	&&\sum_{n=-\infty}^{\infty}e^{iqn\Delta\phi}I_{q|n+\alpha|}(v)=\frac{1}{q}\sum_{k}e^{v\cos(2\pi k/q-\Delta\phi)}e^{i\alpha(2\pi k -q\Delta\phi)}-\frac{e^{-iqn_{0}\Delta\phi}}{2\pi i}\nonumber\\
	&\times&\sum_{j=\pm1}je^{ji\pi q|\alpha_0|} \int_{0}^{\infty}d\eta\frac{(\cosh{[q\eta(1-|\alpha_0|)]}-\cosh{(|\alpha_0| q\eta)e^{-iq(\Delta\phi+j\pi)}})e^{-v\cosh(\eta)}}{\cosh{(q\eta)}-\cos{(q(\Delta\phi+j\pi)}} \ ,
	\label{summation-formula}
\end{eqnarray}
with the parameter $k$ varying  in the interval:
\begin{equation}
	-\frac{q}{2}+\frac{\Delta\phi}{\Phi_{0}}\le k\le \frac{q}{2}+\frac{\Delta\phi}{\Phi_{0}}  \   .
\end{equation}
Substituting \eqref{summation-formula} into \eqref{W-function_R_b}, it is possible to perform the integration over $v$ using \cite{Grad}. The result is
\begin{eqnarray}
	W_{b(R)}(x,x^\prime)&=&-\frac{(ww^\prime)^{D/2}}{2(2\pi)^{D/2}a^{D-1}}\Biggl\{\sum_{k}\frac{e^{i\alpha(2\pi k -q\Delta\phi)}}{u_k^{\frac{D}{2}-1}}\int_{0}^{\infty}dpp^{D/2}\sum_{l=1}^{2}\frac{\bar{J}_{\nu}(pw_0)}{\bar{H}_{\nu}^{(l)}(pw_0)}\nonumber\\
	&\times&H_{\nu}^{(l)}(pw)H_{\nu}^{(l)}(pw^\prime)K_{\frac{D}{2}-1}(pu_k)-\frac{qe^{-iqn_{0}\Delta\phi}}{2\pi i}\sum_{j=\pm1}je^{ji\pi q|\alpha_0|}
	\nonumber\\
	&\times&\int_{0}^{\infty}d\eta\frac{\cosh{[q\eta(1-|\alpha_0|)]}-\cosh{(|\alpha_0| q\eta)e^{-iq(\Delta\phi+j\pi)}}}{u_\eta^{\frac{D}{2}-1}\big[\cosh{(q\eta)}-\cos{(q(\Delta\phi+j\pi))}\big]}\int_{0}^{\infty}dpp^{D/2}\sum_{l=1}^{2}\frac{\bar{J}_{\nu}(pw_0)}{\bar{H}_{\nu}^{(l)}(pw_0)}\nonumber\\
	&\times&H_{\nu}^{(l)}(pw)H_{\nu}^{(l)}(pw^\prime)K_{\frac{D}{2}-1}(pu_\eta)\Biggr\} \ .
	\label{W-function_b-2}
\end{eqnarray}
In the above expression we have introduced the notation
\begin{eqnarray}
	u_{k}^2&=&r^2+r'^2-2rr'\cos{(2\pi k/q-\Delta\phi)}
	+\Delta\vec{x}_{\parallel}^{2}-\Delta t^2\nonumber\\
	u_{\eta}^2&=&r^2+r'^2+2rr'\cosh{(\eta)}+\Delta\vec{x}_{\parallel}^{2}-\Delta t^2.
	\label{arg}
\end{eqnarray}
Finally we rotate the contour integration over $p$ by the angle $\pi/2$ $(-\pi/2)$ for the term $l=1$ $(l=2)$. Using the relations involving Bessel functions with imaginary argument \cite{Abra}, the result is
\begin{eqnarray}
	W_{b(R)}(x,x^\prime)&=&-\frac{(ww^\prime)^{D/2}}{(2\pi)^{D/2}a^{D-1}}\int_{0}^{\infty}dpp^{D-1}\frac{\bar{I}_{\nu}(pw_0)}{\bar{K}_{\nu}(pw_0)}K_{\nu}(pw)K_{\nu}(pw^\prime)\nonumber\\
	&\times&\Biggl\{\sum_{k}e^{i\alpha(2\pi k -q\Delta\phi)}f_{\frac{D}{2}-1}(pu_k)-\frac{qe^{-iqn_{0}\Delta\phi}}{2\pi i}\sum_{j=\pm1}je^{ji\pi q|\alpha_0|}
	\nonumber\\
	&\times&\int_{0}^{\infty}d\eta\frac{\cosh{[q\eta(1-|\alpha_0|)]}-\cosh{(|\alpha_0| q\eta)e^{-iq(\Delta\phi+j\pi)}}}{\cosh{(q\eta)}-\cos{(q(\Delta\phi+j\pi))}}f_{\frac{D}{2}-1}(pu_\eta)\Biggl\} \ ,
	\label{W-function_b-3}
\end{eqnarray}
where we have introduced the notation
\begin{equation}
	f_{\chi}(x)=\frac{J_{\chi}(x)}{x^{\chi}} \ .
	\label{func-f}
\end{equation}

\subsection{$L$-Region}
Now we turn our attention to calculate the positive frequency Wightman function in the $L$-region. Substituting the respective wave function solutions \eqref{Solu-L}, with the coefficient \eqref{C_L-region}, into \eqref{Wightman-def},  we get
\begin{eqnarray}
	W_L(x,x^\prime)&=&\frac{q(ww^\prime)^{D/2}}{a^{D-1}(2\pi)^{D-2}w_0^2}\sum_{\sigma}\frac{\lambda p_{\nu,i}}{\sqrt{(p_{\nu,i}/w_0)^2+\lambda^2+\vec{k}^2}}T_{\nu}(p_{\nu,i})J_{\nu}(p_{\nu,i}w)J_{\nu}(p_{\nu,i}w^\prime)
	\nonumber\\
	&\times&J_{q|n+\alpha|}(\lambda r)J_{q|n+\alpha|}(\lambda r^\prime)e^{inq\Delta\varphi+i\vec{k}\cdot\Delta \vec{x}-iE\Delta t} \ ,
\end{eqnarray}
where now, we have
\begin{equation}
	\sum_{\sigma}=\int_{0}^{\infty}d\lambda\sum_{i=1}^{\infty}\sum_{n}\int d\vec{k} \ .
\end{equation}

Using the identity \eqref{identity}, we can integrate over $\lambda$ and $\vec{k}$ using the formulas of \cite{Grad}, obtaining the following expression:
\begin{eqnarray}
	\label{W_L}
	W_L(x,x^\prime)&=&\frac{q(ww')^{D/2}}{a^{D-1}(2\pi)^{D/2}w_0^2(rr')^{D/2-1}}\int_{0}^{\infty}dvv^{D/2-2}e^{-\frac{r^2+r^{\prime2}+\Delta\vec{x}_{\parallel}^2-\Delta t^2}{2rr^\prime}v}
	\sum_{n=-\infty}^{\infty}e^{inq\Delta\varphi}I_{q|n+\alpha_0|}(v)	\nonumber\\
	&\times&\sum_{i=1}^{\infty}p_{\nu,i}T_{\nu}(p_{\nu,i})J_{\nu}(p_{\nu,i}w/w_0)J_{\nu}(p_{\nu,i}w'/w_0)e^{-\frac{r^2p_{\nu,i}^2}{2w_0^2v}} \ ,
\end{eqnarray}
where we have introduced a new variable $v=r^2/(2s^2)$.

At this point we substitute \eqref{summation-formula} into the above expression. After performing the integral $v$, we obtain,
\begin{eqnarray}
	W_L(x,x^\prime)&=&\frac{2(ww^\prime)^{D/2}}{(2\pi)^{D/2}a^{D-1}w_0^{\frac{D}{2}+1}}\Biggl\{\sum_{k}\frac{e^{i\alpha(2\pi k -q\Delta\phi)}}{u_k^{\frac{D}{2}-1}}\sum_{i=1}^{\infty}p_{\nu,i}^{D/2}T_{\nu}(p_{\nu,i})J_{\nu}(p_{\nu,i}w/w_0)J_{\nu}(p_{\nu,i}w^\prime/w_0)\nonumber\\
	&\times&K_{\frac{D}{2}-1}(u_kp_{\nu,i}/w_0)-\frac{qe^{-iqn_{0}\Delta\phi}}{2\pi i}\sum_{j=\pm1}je^{ji\pi q|\alpha_0|}
	\nonumber\\
	&\times&\int_{0}^{\infty}d\eta\frac{\cosh{[q\eta(1-|\alpha_0|)]}-\cosh{(|\alpha_0| q\eta)e^{-iq(\Delta\phi+j\pi)}}}{u_\eta^{\frac{D}{2}-1}\big[\cosh{(q\eta)}-\cos{(q(\Delta\phi+j\pi))}\big]}\nonumber\\
	&\times&\sum_{i=1}^{\infty}p_{\nu,i}^{D/2}T_{\nu}(p_{\nu,i})J_{\nu}(p_{\nu,i}w/w_0)J_{\nu}(p_{\nu,i}w^\prime/w_0)K_{\frac{D}{2}-1}(u_\eta p_{\nu,i}/w_0)\Biggl\} \ .
	\label{W-function_b-2-L}
\end{eqnarray}
Again, we use the definition \eqref{arg} to the variable $u_k$ and $u_\eta$.

In order to obtain an expression to the Wightman function in the $L$-region more convenient for the extraction of the brane induced part, we apply to the series over $i$ a variant of the generalized Abel-Plana formula \cite{SahaRev}
\begin{equation}
	\sum_{i=1}^{\infty}T_{\nu}(p_{\nu,i})f(p_{\nu,i})=\frac{1}{2}\int_{0}^{\infty}dzf(z)-\frac{1}{2\pi}\int_{0}^{\infty}dz\frac{\bar{K}_{\nu}(z)}{\bar{I}_{\nu}(z)}\left[e^{-i\nu z}f(iz)+e^{i\nu z}f(-iz)\right] \ .
	\label{Abel-Plana}
\end{equation}

For the problem that we are analyzing the function $f(z)$ is given below,
\begin{equation}
	f(z)=z^{D/2}J_{\nu}(zw/w_0)J_{\nu}(zw^\prime/w_0)K_{\frac{D}{2}-1}(2uz/w_0) \ .
	\label{func}
\end{equation}
The first term provided by \eqref{Abel-Plana} corresponds the Wightman function in the absence of brane, while the second is induced by the boundary. As mentioned in the previous subsection, here we are interested in the brane-induced Whigthman function. Therefore, after some intermediate steps, we obtain
\begin{eqnarray}
	W_{b(L)}(x,x^\prime)&=&-\frac{(ww^\prime)^{D/2}}{(2\pi)^{D/2}a^{D-1}}\int_{0}^{\infty}dpp^{D/2}\frac{\bar{K}_{\nu}(pw_0)}{\bar{I}_{\nu}(pw_0)}I_{\nu}(pw)I_{\nu}(pw^\prime)\nonumber\\
	&\times&\Biggl\{\sum_{k}e^{i\alpha(2\pi k -q\Delta\phi)}f_{\frac{D}{2}-1}(pu_k)-\frac{qe^{-iqn_{0}\Delta\phi}}{2\pi i}\sum_{j=\pm1}je^{ji\pi q|\alpha_0|}
	\nonumber\\
	&\times&\int_{0}^{\infty}d\eta\frac{\cosh{[q\eta(1-|\alpha_0|)]}-\cosh{(|\alpha_0| q\eta)e^{-iq(\Delta\phi+j\pi)}}}{\cosh{(q\eta)}-\cos{(q(\Delta\phi+j\pi))}}f_{\frac{D}{2}-1}(pu_\eta)\Biggl\} \ ,
	\label{W-function_b-3-L}
\end{eqnarray}
where we have made change of variable $z=pw_0$.

\section{Bosonic Current Density}
\label{bosonic_cur}
The VEV of the bosonic current density can be expressed in terms of the positive frequency Wightman function by,
	\begin{equation}
	\langle j_{\mu}\rangle=ie\lim_{x^\prime\rightarrow x}\{(\partial_{\mu}-\partial_{\mu^\prime})W(x,x^\prime)+2ieA_{\mu}W(x,x^\prime)\}
	\label{Curr-VEV}
	\end{equation}

Because the analysis of the induced bosonic current	in the $(1+D)-$dimensional AdS space in the presence of a carrying-magnetic-flux cosmic string has been given in \cite{Wagner_19}, here we are mainly interested to calculate the bosonic current induced by the presence of the brane in both regions, i.e., for $0\leq w\leq w_0$ and $w_0\leq w<\infty$. 
	
As we will see in the next subsections, the only non-zero current density components are the azimuthal ones, and it is a periodic function of the magnetic flux along the string, $\Phi_{\phi}$, with period equal to the quantum flux.

\subsection{Charge density}
\label{charge_dens}
Let us begin with the calculation of the charge density. Since $A_{0}=0$, we have
\begin{equation}
	\langle j_{0}(x)\rangle_{b(J)}=ie\lim\limits_{x'\rightarrow x}(\partial_{t}-\partial'_{t})W_{b(J)}(x,x')  \  ,
	\label{charge-density}
\end{equation}
with $J=R,  L$, that represents the $R$ and $L$ regions. 

The analysis of the charge density depends on the behavior of the time derivative of the function $f_{D/2-1}(pu_\sigma)$, with $\sigma=j, \ y$, that appear in \eqref{W-function_b-3} and \eqref{W-function_b-3-L}. Using the fact that
\begin{eqnarray}\partial_zf_\mu(z)=-zf_{\mu+1}(z) \  ,
\end{eqnarray}
and knowing that the arguments of the function $f_{D/2-1}(pu_\sigma)$ depend on the time variable with $(t-t')$,  we can see that
\begin{eqnarray}
\partial_tf_{D/2-1}(pu_\sigma)=p^2(t-t')f_{D/2}(pu_\sigma) \   .
\end{eqnarray} 
Now taking the coincidence limit on the function $f_{D/2}(pu_\sigma)$, we can verify that for the case where $\sigma=k=0$,  this function goes to a finite value for $u_0\to0$. As to the case $\sigma=k\neq0$, $u_k\to2r\sin(k\pi/q)$ and for $\sigma=\eta$, $u_\eta\to 2r\cosh(\eta/2)$. For both last cases the function $f_{D/2}(pu_\sigma)$ assumes a finite value, even for $r\to0$. Finally taking $t'\to t$, the above expression vanishes. Consequently we conclude that there is no induced charge density.

Following similar procedure we can prove that there are no induced current density along the radial coordinate and extra dimensions, i.e., $\langle j^r\rangle=\langle j^i\rangle=0$. As to the induced current along $w$, $\langle j^w\rangle$, we can promptly verify that it is zero.
	
	\section{Azimuthal Current}
	\label{azimuthal_dens}
	In this subsection we will proceed the calculations of the induced azimuthal currents in the regions $R$ and $L$, induced by the brane.
	
\subsection{$R$-region}
	\label{azimuthal_R}
Let us start the analysis in $R$-region. Although the Wightman in this region was given in \eqref{W-function_b-3}, to  develop the calculation of the azimuthal current density, it is more convenient to use the Wightman function provided in \eqref{W-function_R_b}. So, substituting this function into the formal expression for the VEV of the current density operator,
	\begin{eqnarray}
	\langle j_\phi(x)\rangle_{b(R)}=ie\lim_{x'\to x}[(\partial_\phi-\partial_{\phi'})W_{b(R)}(x',x)+2ieA_\phi W_{b(R)}(x',x)] \  ,
	\end{eqnarray}
using $A_{\mu}=\delta_{\mu}^{\phi}A_{\phi}=q\alpha/e$, and taking the coincidence limit also in the angular variable, we get
	
	\begin{eqnarray}
		\langle j_{\phi}(x)\rangle_{b(R)}&=&\frac{qew^{D}}{2a^{D-1}(2\pi)^{D/2}r^{D-2}}\int_{0}^{\infty}{dv}{v^{D/2-2}}e^{-v}\sum_{n}q(n+\alpha)I_{q|n+\alpha|}(v)\nonumber \\
		&\times&\int_{0}^{\infty}dpp e^{-\frac{p^2r^2}{2v}}\sum_{l=1}^{2} \frac{\bar{J}_{\nu}^2(pw_0)}{\bar{H}_{\nu}^{(l)}(pw_0)}\left(H_{\nu}^{(l)}(pw)\right)^2 \ .
		\label{brane}
	\end{eqnarray}

In \cite{Braganca_15}, a compact expression for the summation over the quantum number $n$ above has been derived. We reproduce this result here,
	\begin{eqnarray}
		\sum_{n=-\infty}^{\infty}(n+\alpha)I_{q|n+\alpha|}(v)&=&\frac{2v}{q^2}\sideset{}{'}\sum_{k=1}^{[q/2]}\sin{(2\pi k/q)}\sin{(2\pi k\alpha_0)}e^{v\cos(2\pi k/q)}\nonumber\\
		&+&\frac{v}{q\pi}\int_{0}^{\infty}d\eta\sinh{(\eta)}\frac{e^{-v\cosh{(\eta)}}g(q,\alpha_0,\eta)}{\cosh{(q\eta)}-\cos{(\pi q)}}  \  ,
		\label{Summation-formula}
	\end{eqnarray}
	where $[q/2]$ represents the integer part of $q/2$, and the prime on the sign of
	the summation means that in the case $q=2p$ the term $k=q/2$ should be
	taken with the coefficient $1/2$. Moreover the function, $g(q,\alpha_0,\eta)$, is defined as
	\begin{equation}
		g(q,\alpha_0,\eta)=\sin{(q\pi  \alpha_0)} \sinh{((1-|\alpha_0|)q\eta)}-\sinh{(q\alpha_0 \eta)}\sin{((1-|\alpha_0|)\pi q)}.
		\label{eqn:summation-formula-2}
	\end{equation}
	
	Substituting the above result into \eqref{brane} and with the help of \cite{Grad}, after a few straightforward steps, we get
	\begin{eqnarray}
		\langle j_{\phi}(x)\rangle_{b(R)}&=&\frac{2ew^{D}}{(4\pi)^{D/2}a^{D-1}r^{D/2-2}}\int_{0}^{\infty}dpp^{D/2+1}\Bigg[\sideset{}{'}\sum_{k=1}^{[q/2]}\frac{\sin{(2\pi k/q)}\sin{(2\pi k\alpha_0)}}{s_k^{D/2}}K_{\frac{D}{2}}(2prs_k)
		\nonumber \\
		&+&\frac{q}{2\pi}\int_{0}^{\infty}d\eta\frac{\sinh{(\eta)}g(q,\alpha_0,\eta)}{\cosh^{D/2}(\eta/2)[\cosh(q\eta)-\cos(q\pi)]}K_{\frac{D}{2}}(2pr\cosh(\eta/2))\Bigg]
		\nonumber\\
		&\times&\sum_{l=1}^{2} \frac{\bar{J}_{\nu}^2(pw_0)}{\bar{H}_{\nu}^{(l)}(pw_0)}\left(H_{\nu}^{(l)}(pw)\right)^2 \ , \ s_k=\sin(\pi k/q)  \ .
	\end{eqnarray}
	
As the next step we rotate the contour of the integration over $p$ by the angle $\pi/2$ $(-\pi/2)$ for the term with $l=1$ $(l=2)$. Using the relations between Bessel functions of imaginary argument, the contravariant component of the azimuthal current reads
	\begin{eqnarray}
		\label{azimu_brane_R}
		\langle j^{\phi}\rangle_{b(R)}&=&-\frac{4e}{(2\pi)^{D/2}a^{D+1}}\int_{0}^{\infty}dzz^{D+1}\frac{\bar{I}_{\nu}(zw_0/w)}{\bar{K}_{\nu}(zw_0/w)}K_{\nu}^2(z)\nonumber\\
		&\times&\Bigg[\sideset{}{'}\sum_{k=1}^{[q/2]}\sin{(2\pi k/q)}\sin{(2\pi k\alpha_0)}f_{\frac{D}{2}}(2z(r/w)s_k)
		\nonumber \\
		&+&\frac{q}{2\pi}\int_{0}^{\infty}d\eta\frac{\sinh{(\eta)}g(q,\alpha_0,\eta)}{\cosh(q\eta)-\cos(q\pi)}f_{\frac{D}{2}}(2z(r/w)\cosh(\eta/2))\Bigg] \ ,
	\end{eqnarray}
where we have defined $z=p w$. Note that this VEV depends on the
ratio $r/w$, which is related to the proper distance from the string, and the ratio $w/w_0$, which is related to the proper distance from the brane $w/w_0=e^{(y-y_0)/a}$.

Let us now investigate this VEV in some special and asymptotic cases. In the conformal massless field case, we have $\nu=1/2$ according to \eqref{nu}, and using the corresponding expressions of the Bessel functions for this particular order, we obtain
\begin{eqnarray}
	\label{azimu_brane_R_m0}
	\langle j^{\phi}\rangle_{b(R)}&=&-\frac{4e}{(2\pi)^{D/2}a^{D+1}}\int_{0}^{\infty}dzz^{D}e^{(-2+\frac{w_0}{w})z}\nonumber\\
	&\times&\frac{2B_0(w_0/w)z\cosh(zw_0/w)+(2A_0-B_0)\sinh(zw_0/w)}{2A_0-B_0(1+2zw_0/w)}\nonumber\\
	&\times&\Bigg[\sideset{}{'}\sum_{k=1}^{[q/2]}\sin{(2\pi k/q)}\sin{(2\pi k\alpha_0)}f_{\frac{D}{2}}(2z(r/w)s_k)
	\nonumber \\
	&+&\frac{q}{2\pi}\int_{0}^{\infty}d\eta\frac{\sinh{(\eta)}g(q,\alpha_0,\eta)}{\cosh(q\eta)-\cos(q\pi)}f_{\frac{D}{2}}(2z(r/w)\cosh(\eta/2))\Bigg] \ .
\end{eqnarray}

For distances from the brane much larger compared with the AdS
radius, $w/w_0\gg1$, we make use of the formulae
for the modified Bessel functions for small values of the argument \cite{Abra}, with the assumption that
$A_0-\nu B_0\neq0$, to the leading order, we have
\begin{eqnarray}
	\label{azimu_brane_R-asymp}
	\langle j^{\phi}\rangle_{b(R)}&\approx&-\frac{2^{3-2\nu-D/2}e}{\pi^{D/2}\Gamma(\nu)\Gamma(\nu+1)a^{D+1}}\left(\frac{A_0+\nu B_0}{A_0-\nu B_0}\right)\left(\frac{w_0}{w}\right)^{2\nu}\int_{0}^{\infty}dzz^{D+2\nu+1}K_{\nu}^2(z)\nonumber\\
	&\times&\Bigg[\sideset{}{'}\sum_{k=1}^{[q/2]}\sin{(2\pi k/q)}\sin{(2\pi k\alpha_0)}f_{\frac{D}{2}}(2z(r/w)s_k)
	\nonumber \\
	&+&\frac{q}{2\pi}\int_{0}^{\infty}d\eta\frac{\sinh{(\eta)}g(q,\alpha_0,\eta)}{\cosh(q\eta)-\cos(q\pi)}f_{\frac{D}{2}}(2z(r/w)\cosh(\eta/2))\Bigg] \ .
\end{eqnarray}

Another interesting limiting case is the Minkowskian limit. In this asymptotic limit, we take $a\rightarrow\infty$ with fixed $y$, and the geometry under consideration is reduced to the background of a cosmic string in $(D+1)$-dimensional Minkowski spacetime. As we approach the Minkowskian limit, the coordinate $w$ in the arguments of the Bessel functions become large and one has $w\approx a+y$. By taking into account that also the order becomes large in this limit, we can use the corresponding uniform asymptotic expansion of the Bessel function in \eqref{azimu_brane_R}. After a few intermediate  steps, to the leading order, we obtain
\begin{eqnarray}
	\label{azimu_brane_R-M_Asymptotic}
	\langle j^{\phi}\rangle_{b}^{(\rm{M})}&=&-\frac{2e}{(2\pi)^{D/2}}\int_{m}^{\infty}du(u^2-m^2)^{D/2}\Bigg[\sideset{}{'}\sum_{k=1}^{[q/2]}\sin{(2\pi k/q)}\sin{(2\pi k\alpha_0)}\nonumber\\
	&\times&f_{\frac{D}{2}}(2rs_k\sqrt{u^2-m^2})
	+\frac{q}{2\pi}\int_{0}^{\infty}d\eta\frac{\sinh{(\eta)}g(q,\alpha_0,\eta)}{\cosh(q\eta)-\cos(q\pi)}\nonumber\\
	&\times&f_{\frac{D}{2}}(2r\cosh(\eta/2)\sqrt{u^2-m^2})\Bigg]\frac{1+\beta u}{1-\beta u}e^{-2u(y-y_0)} \ .
\end{eqnarray}
For the Neumann BC, $\beta\rightarrow\infty$, we can perform the last integration by making the change of variable $v=\sqrt{u^2-m^2}$ and using the identity \eqref{identity}, the expression given in \eqref{azimu_brane_L-M_Asymptotic} is reduced to
\begin{eqnarray}
	\label{azimu_brane_R-M}
	\langle j^{\phi}\rangle_{b}^{(\rm{M})}&=&\frac{4em^{D+1}}{(2\pi)^{(D+1)/2}}\Bigg[\sideset{}{'}\sum_{k=1}^{[q/2]}\sin{(2\pi k/q)}\sin{(2\pi k\alpha_0)}\nonumber\\
	&\times&\frac{K_{\frac{D+1}{2}}\left(2m[r^2s_k^2+(y-y_0)^2]\right)}{\left(2m[r^2s_k^2+(y-y_0)^2]\right)^{\frac{D+1}{2}}}
	+\frac{q}{2\pi}\int_{0}^{\infty}d\eta\frac{\sinh{(\eta)}g(q,\alpha_0,\eta)}{\cosh(q\eta)-\cos(q\pi)}\nonumber\\
	&\times&\frac{K_{\frac{D+1}{2}}\left(2m[r^2\cosh^2(\eta/2)+(y-y_0)^2]\right)}{\left(2m[r^2\cosh^2(\eta/2)+(y-y_0)^2]\right)^{\frac{D+1}{2}}}\Bigg] \ ,
\end{eqnarray}
which coincides exactly with the result reported in \cite{Braganca:2020jci} for $y_0=0$. On the other hand, for Dirichlet BC, $\beta=0$, the corresponding result differs from \eqref{azimu_brane_R-M} by the sign.

\begin{figure}[!htb]
	\begin{center}
		\includegraphics[scale=0.3]{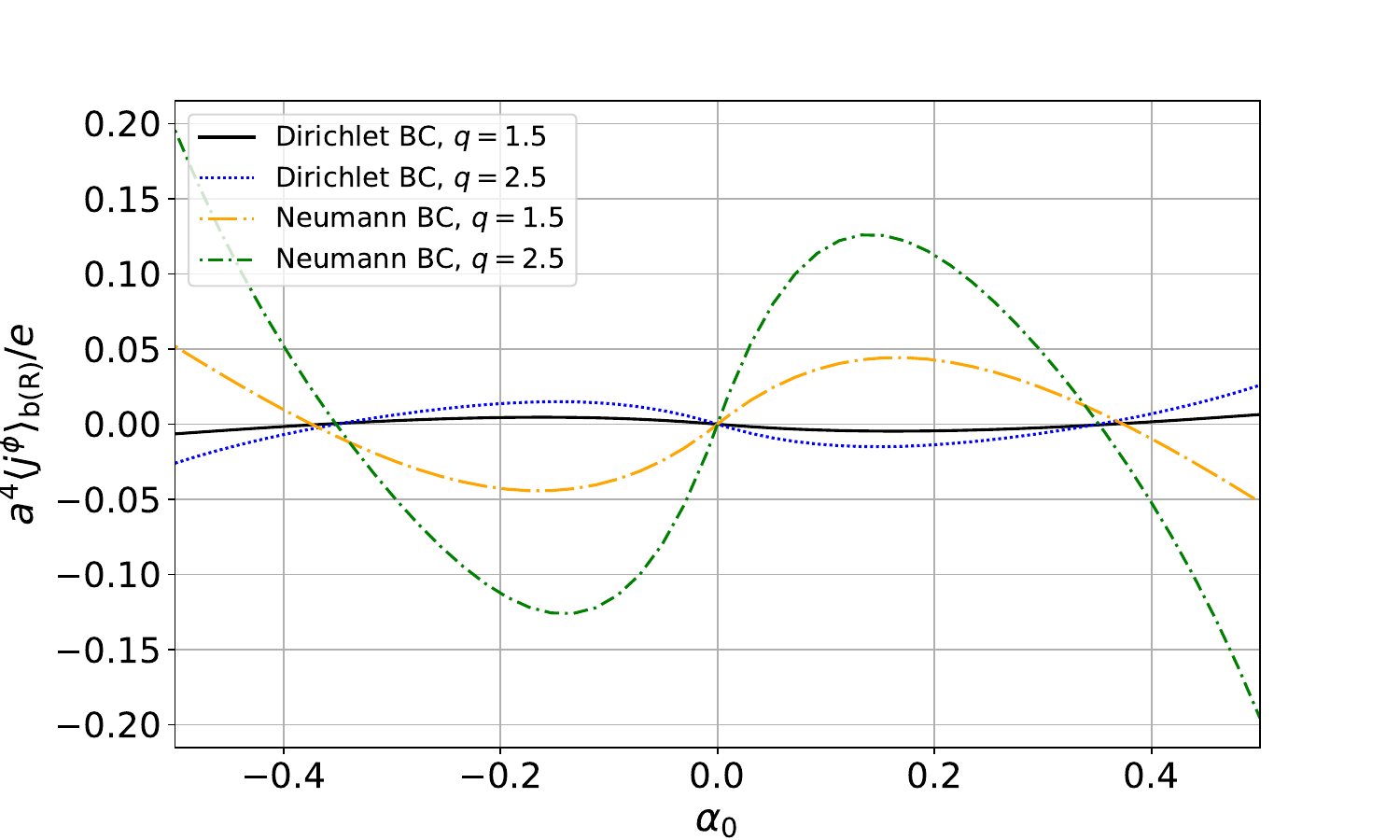}
		\quad
		\includegraphics[scale=0.3]{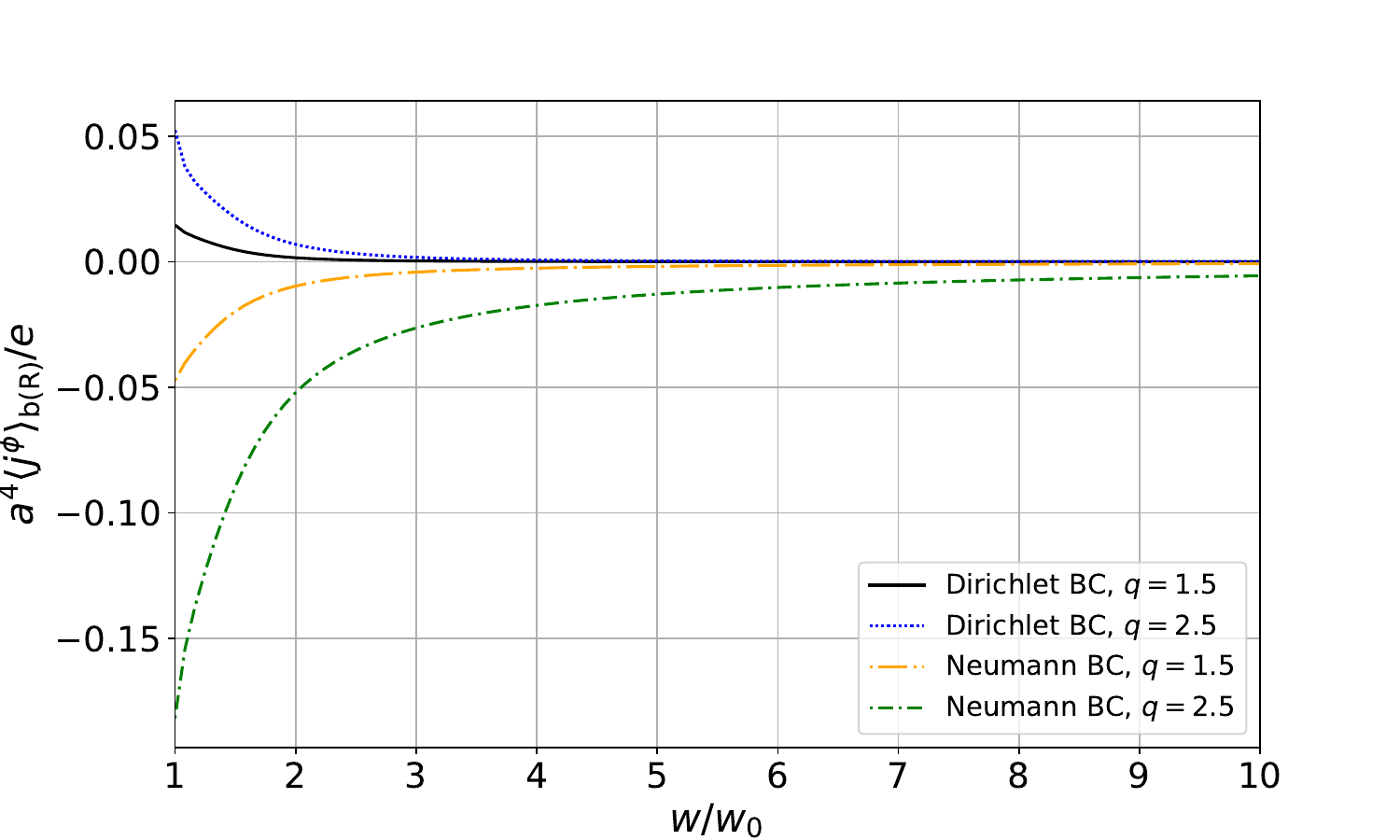}
		\caption{The VEV of the current density in the $R$-region for $D=3$ is plotted as function of the magnetic flux, $\alpha_0$, (left panel) and the proper distance from the brane in units of $a$, $w/w_0$, considering Dirichlet and Neumann boundary conditions and different values of $q$. Both graphs are plotted for $D=3$ and $r/w=0.5$. Moreover, in the left panel we have fixed $w/w_0=2$ and in the right one, $\alpha_0=0.4$.}
		\label{fig1}
	\end{center}
\end{figure}

In Fig. \ref{fig1} is exhibit the behavior of the current density in $R$-region as function of the magnetic flux, $\alpha_0$, (left panel) and the ratio $w/w_0$ (right panel), considering Dirichlet and Neumann boundary conditions with different values of the parameter associated with the deficit angle, $q$. From the left panel we can see that the current density is an odd function of $\alpha_0$, as expected. From the right panel we observe that the VEV is finite on the brane location and goes to zero as it approaches the AdS horizon, in accordance to our asymptotic analysis. Moreover, note that in both plots the intensities increase with $q$ and are higher for Neumann BC.
\subsection{$L$-region}
\label{azimuthal_L}

To calculate the induced azimuthal current density in the $L$-region, we start by substituting the Wightman function given in \eqref{W_L} into the expression bellow,
	\begin{eqnarray}
		\label{Curr_L-region}
	\langle j_\phi(x)\rangle_{b(L)}=ie\lim_{x'\to x}[(\partial_\phi-\partial_{\phi'})W_{b(L)}(x',x)+2ieA_\phi W_{b(L)}(x',x)] \ ,
\end{eqnarray}
using $A_{\mu}=\delta_{\mu}^{\phi}A_{\phi}=q\alpha/e$. After some steps we get,
\begin{eqnarray}
	\label{curr_L_brane}
	\langle j^{\phi}\rangle_{b(L)}&=&-\frac{2eq^2w^{D}}{a^{D-1}(2\pi)^{D/2}w_0^2r^{D-2}}\int_{0}^{\infty}dvv^{D/2-2}e^{-v} \sum_{n=-\infty}^{\infty}(n+\alpha)I_{q|n+\alpha_0|}(v)
	\nonumber\\
	&\times&\sum_{i=1}^{\infty}p_{\nu,i}T_{\nu}(p_{\nu,i})J_{\nu}^2(p_{\nu,i}w/w_0)e^{-\frac{r^2p_{\nu,i}^2}{2w_0^2v}} \ .
\end{eqnarray}
Substituting the summation over $n$ in the modified Bessel function by \eqref{Summation-formula}, we have,
\begin{eqnarray}
	\label{curr_L_brane_1}
	\langle j_{\phi}\rangle_{b(L)}&=&-\frac{4ew^{D}r^{2-D}}{a^{D-1}(2\pi)^{D/2}w_0^2}\int_{0}^{\infty}dvv^{D/2-1}e^{-v} \left[\sideset{}{'}\sum_{k=1}^{[q/2]}\sin{(2\pi k/q)}\sin{(2\pi k\alpha_0)}e^{v\cos(2\pi k/q)}\right.\nonumber\\
	&+&\left.\frac{q}{2\pi}\int_{0}^{\infty}d\eta\sinh{(\eta)}\frac{e^{-v\cosh{(\eta)}}g(q,\alpha_0,\eta)}{\cosh{(q\eta)}-\cos{(\pi q)}}\right]  
\sum_{i=1}^{\infty}p_{\nu,i}T_{\nu}(p_{\nu,i})J_{\nu}^2(p_{\nu,i}w/w_0)e^{-\frac{r^2p_{\nu,i}^2}{2w_0^2v}} .
\end{eqnarray}
Now we are in position to integrate over the variable $v$. Doing this we get,
\begin{eqnarray}
	\label{curr_L_brane_2}
	\langle j_{\phi}\rangle_{b(L)}&=&-\frac{8ew^{D}}{a^{D-1}(4\pi)^{D/2}r^{D/2-2}w_0^{D/2+2}}	\sum_{i=1}^{\infty}p^{D/2+1}_{\nu,i}T_{\nu}(p_{\nu,i})J_{\nu}^2(p_{\nu,i}w/w_0)\nonumber\\
	&\times&\left[\sideset{}{'}\sum_{k=1}^{[q/2]}\frac{\sin{(2\pi k/q)}\sin{(2\pi k\alpha_0)}}{s_k^{D/2}}K_{\frac{D}{2}}(2s_k r p_{\nu,i}/w_0)\right.\nonumber\\
	&+&\left.\frac{q}{2\pi}\int_{0}^{\infty}d\eta\frac{\sinh{(\eta)}g(q,\alpha_0,\eta)}{\cosh^{D/2}(\eta/2)[\cosh{(q\eta)}-\cos{(\pi q)}]}K_{\frac{D}{2}}(2\cosh(\eta/2) r p_{\nu,i}/w_0)\right]  \ .
\end{eqnarray} 

Now to proceed the summation over the quantum number $i$, we use the generalized Abel-Plana summation formula  \eqref{Abel-Plana}, with 
\begin{eqnarray}
	f(z)=z^{D/2+1}J^2_\nu(z(w/w_0))K_{\frac{D}{2}}(2 \rho z(r/w_0)) \ , 
\end{eqnarray}
with $\rho$ being $\sin(\pi k/q)$ or $\cosh(\eta/2)$. 

Because we want to calculate the azimuthal current induced by the brane, only the second term on the right hand side of the summation formula is of our interest. Using the relations between the Bessel functions with imaginary arguments \cite{Abra}, and after some intermediate steps, we obtain
\begin{eqnarray}
	\langle j_{\phi}\rangle_{b(L)}&=&\frac{4ew^D}{(4\pi)^{D/2}a^{D-1}}\frac1{w_0^{D/2+2}r^{D/2-2}} \int_{0}^{\infty}dzz^{D/2+1}\frac{\bar{K}_{\nu}(z)}{\bar{I}_{\nu}(z)}I_{\nu}^2(z(w/w_0))\nonumber\\
	&\times&\Bigg[\sideset{}{'}\sum_{k=1}^{[q/2]}\frac{\sin{(2\pi k/q)}\sin{(2\pi k\alpha_0)}}{s_k^{D/2}}J_{\frac{D}{2}}(2z(r/w_0)s_k)
	\nonumber \\
	&+&\frac{q}{2\pi}\int_{0}^{\infty}d\eta\frac{\sinh{(\eta)}g(q,\alpha_0,\eta)}{\cosh^{D/2}(\eta/2)[\cosh(q\eta)-\cos(q\pi)]}J_{\frac{D}{2}}(2z(r/w_0)\cosh(\eta/2))\Bigg] \ .
\end{eqnarray}
Now changing the variable $z\to z(w_0/w)$, the contravariant component of the azimuthal current reads
\begin{eqnarray}
	\label{azimu_brane_L}
	\langle j^{\phi}\rangle_{b(L)}&=&-\frac{4e}{(2\pi)^{D/2}a^{D+1}}\int_{0}^{\infty}dzz^{D+1}\frac{\bar{K}_{\nu}(zw_0/w)}{\bar{I}_{\nu}(zw_0/w)}I_{\nu}^2(z)\nonumber\\
	&\times&\Bigg[\sideset{}{'}\sum_{k=1}^{[q/2]}\sin{(2\pi k/q)}\sin{(2\pi k\alpha_0)}f_{\frac{D}{2}}(2z(r/w)s_k)
	\nonumber \\
	&+&\frac{q}{2\pi}\int_{0}^{\infty}d\eta\frac{\sinh{(\eta)}g(q,\alpha_0,\eta)}{\cosh(q\eta)-\cos(q\pi)}f_{\frac{D}{2}}(2z(r/w)\cosh(\eta/2))\Bigg] \ .
\end{eqnarray}
Comparing the above expression  with \eqref{azimu_brane_R}, we observe that the brane-induced azimuthal current in the $L$-region is obtained from the corresponding one in $R$-region by the replacements $I\rightarrow K$,
$K\rightarrow I$ of the modified Bessel functions.

Proceeding similarly as in the $R$-region, we now investigate some special and limiting cases. In the conformal massless field case, with $\nu=1/2$, we use the corresponding expressions of the Bessel functions, and after a few algebraic manipulations, we get
\begin{eqnarray}
	\label{azimu_brane_L-m0}
	\langle j^{\phi}\rangle_{b(L)}&=&-\frac{4e}{(2\pi)^{D/2}a^{D+1}}\int_{0}^{\infty}dzz^{D}e^{-zw_0/w}\nonumber\\
	&\times&\frac{[2A_0-B_0(1+2zw_0/w)]\sinh^2(z)}{2B_0(w_0/w)z\cosh(zw_0/w)+(2A_0-B_0)\sinh(zw_0/w)}I_{\nu}^2(z)\nonumber\\
	&\times&\Bigg[\sideset{}{'}\sum_{k=1}^{[q/2]}\sin{(2\pi k/q)}\sin{(2\pi k\alpha_0)}f_{\frac{D}{2}}(2z(r/w)s_k)
	\nonumber \\
	&+&\frac{q}{2\pi}\int_{0}^{\infty}d\eta\frac{\sinh{(\eta)}g(q,\alpha_0,\eta)}{\cosh(q\eta)-\cos(q\pi)}f_{\frac{D}{2}}(2z(r/w)\cosh(\eta/2))\Bigg] \ .
\end{eqnarray}

For points near the AdS boundary ($w = 0$), $w\ll w_0$, the argument of the modified Bessel function $I_{\nu}(z)$ is small and using the corresponding asymptotic function in the leading term \cite{Abra}, we get
\begin{eqnarray}
	\label{azimu_brane_L-asymp}
	\langle j^{\phi}\rangle_{b(L)}&\approx&-\frac{2^{2-2\nu-D/2}e}{\pi^{D/2}\Gamma^2(\nu+1)a^{D+1}}\left(\frac{w}{w_0}\right)^{2\nu}\int_{0}^{\infty}dzz^{D+2\nu+1}\frac{\bar{K}_{\nu}(zw_0/w)}{\bar{I}_{\nu}(zw_0/w)}\nonumber\\
	&\times&\Bigg[\sideset{}{'}\sum_{k=1}^{[q/2]}\sin{(2\pi k/q)}\sin{(2\pi k\alpha_0)}f_{\frac{D}{2}}(2z(r/w)s_k)
	\nonumber \\
	&+&\frac{q}{2\pi}\int_{0}^{\infty}d\eta\frac{\sinh{(\eta)}g(q,\alpha_0,\eta)}{\cosh(q\eta)-\cos(q\pi)}f_{\frac{D}{2}}(2z(r/w)\cosh(\eta/2))\Bigg] \ .
\end{eqnarray}

In the Minkowskian limit, we follow the same procedure described for the current density in the $R$-region, obtaining the following result:
\begin{eqnarray}
	\label{azimu_brane_L-M_Asymptotic}
	\langle j^{\phi}\rangle_{b(L)}^{(\rm{M})}&=&-\frac{2e}{(2\pi)^{D/2}}\int_{m}^{\infty}du(u^2-m^2)^{D/2}\Bigg[\sideset{}{'}\sum_{k=1}^{[q/2]}\sin{(2\pi k/q)}\sin{(2\pi k\alpha_0)}\nonumber\\
	&\times&f_{\frac{D}{2}}(2rs_k\sqrt{u^2-m^2})
	+\frac{q}{2\pi}\int_{0}^{\infty}d\eta\frac{\sinh{(\eta)}g(q,\alpha_0,\eta)}{\cosh(q\eta)-\cos(q\pi)}\nonumber\\
	&\times&f_{\frac{D}{2}}(2r\cosh(\eta/2)\sqrt{u^2-m^2})\Bigg]\frac{1+\beta u}{1-\beta u}e^{-2u(y_0-y)} \ ,
\end{eqnarray}
which is similar to the result we found for the $R$-region with $y-y_0$ replaced by $y_0-y$. This similarity is expected since in the Minkowskian limit the VEV is symmetric to the brane. Moreover, for Neumann BC we obtain the same result present in \eqref{azimu_brane_R-M} with $y-y_0$ replaced by $y_0-y$. On the other hand, for Dirichlet BC  the corresponding expression differs by the sign. 

\begin{figure}[!htb]
	\begin{center}
		\includegraphics[scale=0.3]{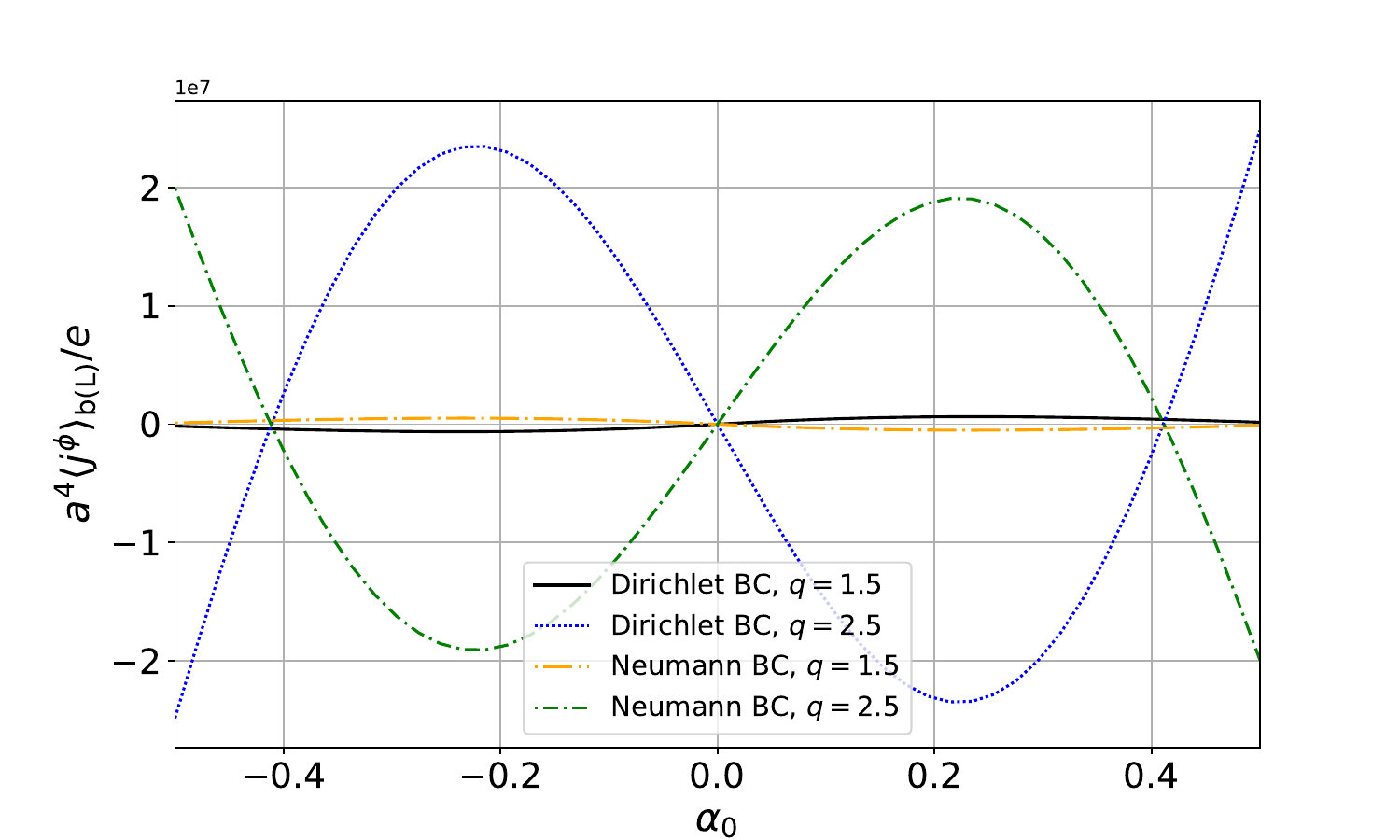}
		\quad
		\includegraphics[scale=0.3]{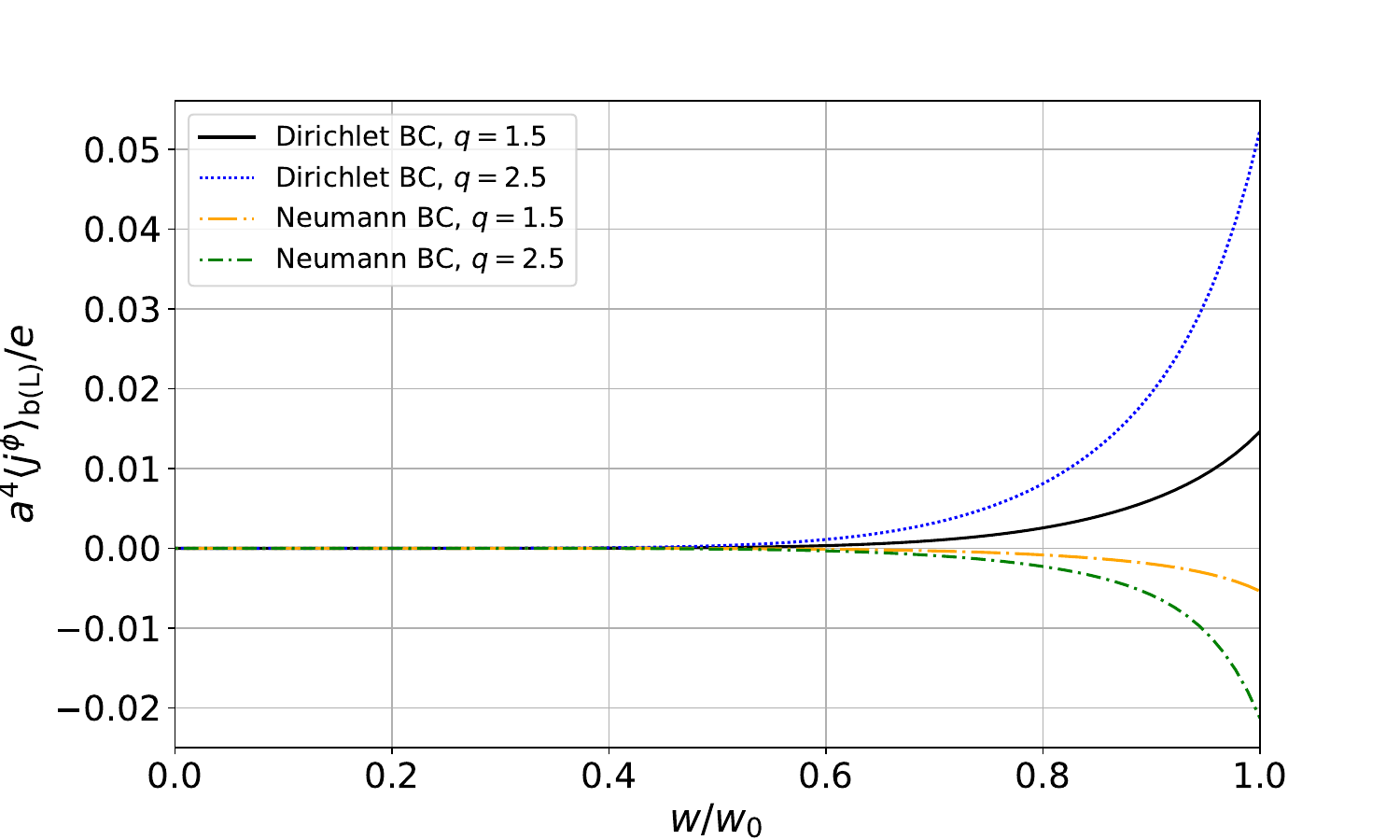}
		\caption{The same as in Fig. \ref{fig1} for the $L$-region.}
		\label{fig2}
	\end{center}
\end{figure}

In Fig. \ref{fig2} is displayed the dependence of the current density in the $L$-region as function of the magnetic flux, $\alpha_0$, (left panel) and the ratio $w/w_0$ (right panel), considering Dirichlet and Neumann boundary conditions with different values of the parameter associated with the deficit angle, $q$. Similar to the $R$-region, from the left panel we observe that the current density in the $L$-region is an odd function of $\alpha_0$. On the other hand, from the right panel we can see that the VEV rapidly goes to zero near the AdS boundary and is finite on the brane. Moreover, in both plots the intensities increase with the parameter $q$ and, differently from the $R$-region, are higher for Dirichlet BC.
\section{Application to RSII model}
\label{RS}
The formulas for the VEV of the current density in the setup problem under consideration and present in the previous section can be applied to the study of the influence of the cosmic string in $Z_2$-symmetric braneworlds models. In this paper we apply them to the Randall-Sundrum model with a single brane (RSII) \cite{RSII}. In this model, the four-dimensional world is understood as a  $Z_2$-symmetric brane with positive tension embedded in a higher dimensional AdS bulk. Moreover the existence of bulk fields is a common feature in models motivated by string theories. In the present setup problem, we consider the cosmic string perpendicular to the brane. The latter is located at $y=0$ (or $w_0=a$ in the $w$ coordinate) and divides the background geometry in two copies of the $R$-region which are identified by the $Z_2$-symmetry  transformation $y\longleftrightarrow -y$. The corresponding line element is obtained from \eqref{ds1} by the replacement $e^{-2y/a}\rightarrow e^{-2|y|/a}$. The fields in the regions $-\infty<y<0$ and $0<y<\infty$ are related by the $Z_2$-symmetry of the model. It worths to note that for an observer located at $y=0$, the corresponding line element for the described geometry is reduced to the one for a cosmic string in $(D+1)$-dimensional flat spacetime. 

The bulk scalar field obeys the boundary conditions at the position of the brane and are obtained by integration of the field equations about $y=0$ \cite{Saharian:2003qs,Gherghetta2000,Flachi2001}. Following this procedure for an untwisted scalar field (even under reflection with respect to the brane location), one can see that the received boundary condition has the Robin form \eqref{RBC}, with coefficient $\beta=a/4D\xi$. In the particular case of a minimally coupled field, $\xi=0$, the boundary condition is reduced to the Neumann one ($\beta\rightarrow\infty$). On the other hand, for twisted scalar fields (odd under reflection with respect to the brane), the boundary condition has the Dirichlet form.

In models with $Z_2$ symmetry, the range of integration over the extra dimension $y$ varies from $-\infty$ to $\infty$ in the normalization condition. This results in an additional factor of $1/2$  when compared to the one obtained previously for the $R$-region, which has half of interval range, $0\le y<\infty$, with the brane location at $y_0=0$. Thus, we can conclude that, similar to the VEVs of the field squared and the energy-momentum tensor \cite{Wagner_23}, the formulas for the VEV of the current density induced by a cosmic string in the generalized RSII braneworld model are obtained from those expressions present in subsection \ref{azimuthal_R} by directly putting $w_0=a$ with an additional overall factor of $1/2$.
	
\section{Conclusions}
\label{conc} 
In the present paper we have studied the combined effects of curvature, conical topology and the presence of a brane on the VEV of the current density for a massive charged scalar field propagating in the background of a $(D+1)$-dimensional AdS spacetime. Along the cosmic string we assume the existence of a magnetic flux. Moreover, on the brane we impose that the field operator obeys the general Robin boundary conditions, and we consider that the string is perpendicular to the brane which is parallel to the AdS boundary, which divides the manifold into two regions, $L$-region ($0\le w\le w_0$) and $R$-region ($w_0\le w<\infty)$. In this setup the scalar modes are obtained for both regions and the corresponding Wightman functions. They are presented in closed form in Section \ref{Wight_fun} for the $R$-region \eqref{W-function_b-3} and $L$-region \eqref{W-function_b-3-L}. 

In Section \ref{bosonic_cur} we have calculated the VEVs of the current density for the $R$-region and $L$-region. We have shown that the only non-vanishing component is induced along the azimuthal direction. In Subsection \ref{azimuthal_R} we have developed the VEV of the azimuthal current density in the $R$-region and a closed expression is present in \eqref{azimu_brane_R}. For a massless quantum scalar field, the azimuthal current density is given by \eqref{azimu_brane_R_m0} for general Robin boundary conditions. For distances from the brane much larger compared with the AdS radius, $w/w_0\gg1$, we found that the $\langle j^{\phi}\rangle_{b(R)}$ decays as $(w_0/w)^{2\nu}$. Also we have obtained its Minkowskian limit given by the  expression \eqref{azimu_brane_R-M_Asymptotic} for the general Robin boundary condition. For the particular cases of Neumann and Dirichet boundary condition we found that the VEV in this limiting case differs by the sign. In order to provide a better understanding of this induced current, in figure \ref{fig1} we exhibit the plots for behaviors of the azimuthal current density, considering $D=3$ and fixed $r/w$, in the $R$-region, $\langle j^{\phi}\rangle_{b(R)}$, as function of the magnetic flux along the string, $\alpha_0$, (left panel) and the ratio $w/w_0$ (right panel). The left panel shows that this VEV is an odd function of the magnetic flux. The right panel shows that $\langle j^{\phi}\rangle_{b(R)}$ is finite on the brane and goes to zero as it approaches the AdS horizon, confirming the asymptotic analysis. In addition, both plots show that the intensities increase with parameter associated with deficit angle, $q$, and are higher for Neumann BC.

In Subsection \ref{azimuthal_L} we have developed the azimuthal current density induced in the $L$-region. The corresponding formula in closed form is given by \eqref{azimu_brane_L} and is obtained from the result for $R$-region by the replacement $I\rightarrow K$, $K\rightarrow I$ of the Bessel functions. For a conformal massless scalar field, $\langle j^{\phi}\rangle_{b(L)}$ is present in \eqref{azimu_brane_L-m0}. For points close to the AdS boundary, $w\ll w_0$, we found the expression \eqref{azimu_brane_L-asymp} which shows that the azimuthal current density in this region decays as $(w/w_0)^{2\nu}$. In the Minkowskian limit we have obtained the expression present in \eqref{azimu_brane_L-M_Asymptotic}, which is similar to the result we found in the $R$-region with $y-y_0$ replaced by $y_0-y$. Moreover, in the particular cases of Neumann and Dirichlet boundary conditions, the VEV in this limiting case differs by the sign. For further investigation, in figure \ref{fig2} we have plotted the azimuthal current density given by \eqref{azimu_brane_L} as function of $\alpha_0$ (left panel) and $w/w_0$ (right panel), considering $D=3$ and fixed $r/w$. Similar to the $R$-region, the left panel shows that $\langle j^{\phi}\rangle_{b(L)}$ is an odd function of the magnetic flux long the string, $\alpha_0$, as expected. The right panel shows that azimuthal current density rapidly vanishes near the AdS boundary, in accordance to the asymptotic analysis, and is finite on the brane. Moreover, both plots show that the intensities of the azimuthal current density increase with the parameter $q$ and, in contrast with the $R$-region, are higher for Dirichlet BC. 

In the last section we have applied the results found in the $R$-region
to investigate the cosmic string induced effects in the generalized
RSII model. By integration of
the field equations about the brane location, $y=0$, we have found that the boundary
conditions in this $Z_2$-symmetric model has the Robin type
with the coefficient $\beta=a/4D\xi$ for untwisted scalar field and it is reduced to the Neumann BC for a minimally coupled field,
$\xi=0$. On the other hand
for a twisted scalar field, we have the Dirichlet BC. As a final remark, we have concluded that the
VEV of the azimuthal current density
induced by a cosmic string in the RSII
model can be obtained from those present in the Section \ref{RS} by putting $w_0 = a$ with an additional factor 1/2.
\section*{Acknowledgment}
W.O.S is supported under grant 2022/2008, Paraíba State Research Foundation (FAPESQ). E.R.B.M is partially supported by CNPq under Grant no 301.783/2019-3.

\end{document}